\documentclass[pra,10pt,tightenlines,twoside,twocolumn,showpacs]{revtex4}

\usepackage{graphicx}
\usepackage{color}
\usepackage{epsfig}
\usepackage{amsmath}
\usepackage{amstext}
\usepackage{amssymb}
\usepackage{bm}
\usepackage[]{hyperref}

\begin{document}
\author{Mateusz Cholascinski}
\affiliation{Nonlinear Optics Division, Adam Mickiewicz University,
  61614 Poznan, Poland} 
\affiliation{Institut f\"ur Theoretische Festk\"orperphysik,
Universit\"at Karlsruhe, D-76128 Karlsruhe, Germany}

\date{\today}
\title{Geometry of an adiabatic passage at a
  level crossing}

\begin{abstract}
We discuss adiabatic quantum phenomena at a level crossing. Given a
path in the parameter space which passes through a degeneracy point,
we find a criterion which determines whether the adiabaticity condition
can be satisfied. For paths that can be traversed adiabatically we
also derive a differential equation which specifies the time dependence
of the system parameters, for which transitions between distinct
energy levels can be neglected. We also generalize the well-known
geometric connections to the case of adiabatic paths containing
arbitrarily many level-crossing points and degenerate levels.
\end{abstract}

\pacs{32.80.Bx, 03.65.Vf}

\maketitle

\section{Introduction}\label{intro}

Evolution of a quantum system, if governed by a time-independent
Hamiltonian, is fully described by relative phase shifts acquired by
its energy eigenstates. These phase shifts which we refer to as
dynamical phases are equal to the energy differences of the levels
integrated over time. On the other hand, for a general time-dependent
system Hamiltonian fully analytical treatment is usually
impossible, or at least hardly tractable. There is also an
intermediate regime in which adiabatic 
variation of the system parameters leaves the population of the
instantaneous energy levels unchanged. In this case, however, apart
from the dynamical phases the levels acquire geometric phase factors
\cite{berry,samuel,panchara} or, in the case of degeneracy, undergo a
nontrivial transformation (within the degenerate subspace), also
geometric in its nature \cite{WZ}. A separate class of
phenomena utilizes only the process of adiabatic following (independently
of the phases acquired), which can result in a coherent population
transfer between levels that are not directly coupled [this process
has been extensively studied in quantum optical systems, where is
known as the stimulated Raman adiabatic passage
(STIRAP)\cite{bergmann}]. The peculiar property of the latter is that
the entire transformation takes place within a 
single, non-degenerate level, and the resulting transformation is not
simply just a phase shift. This seems to contradict the predictions
made by Berry \cite{berry}. However, more detailed study shows that such
transfer is possible because of the level crossings at the initial and
final times of the process. 

The theory of adiabatic phenomena is well-established for systems with
exactly distinguishable or exactly degenerate levels. The vicinity of
the crossing points seems to be a troublesome, intermediate
regime: when the energy difference between the energy levels
tends to zero there is apparently no time scale defining the
adiabaticity. Recently, \textcite{avron} discussed the adiabatic theorem 
in dissipative systems. In this scenario the energy gap, which is present in
isolated systems, disappears. Nevertheless, as the authors show, the
adiabaticity still can be defined. Here we consider an isolated quantum
system in which the levels cross as the externally controlled
parameters are varied. An analysis of adiabatic phenomena in isolated
systems with nearly crossing levels has been performed
in Ref.~\cite{low}; however, the time dependence of the system
parameters has been assumed to be arbitrary. Similarly, in Ref.~\cite{stenholm}
the nonadiabatic corrections due to level crossing in three-level
processes (STIRAP) has been evaluated, again assuming that the
time dependence of the parameters is fixed for a given
realization. In our work we perform an analysis of an evolution in 
which only the path in the parameter space is arbitrary. It is
indeed very often 
the most important characteristic in experiments probing adiabatic
phenomena, such as detection of geometric phases or
population transfer (STIRAP, coherent charge pumping
\cite{pothier,pekola1,keller}), and the time dependence does not
influence the results of the measurements any further (it needs only to
be adiabatic).

To summarize the most important of our results, let us suppose that we
have given a path in the 
parameter space that passes through some degeneracy point and we want
to determine whether it is possible to satisfy the adiabaticity condition. 
In what follows, we show that the answer is uniquely determined by the
geometry of the path. For the cases when it is possible, we derive a
differential equation whose 
solution gives the {\em time} variation of the parameters (along the
path), for which transitions between distinct energy levels can
be neglected. 
Moreover, we show that at the level crossing the energy
eigenstates are discontinuous, which can result in  nontrivial
transformations. The points of crossing correspond themselves to
geometric phenomena, in which the geometry is determined only by
the direction in the parameter space from which such points are
approached. For closed paths which can be passed adiabatically our
results together with the Wilczek-Zee connection \cite{WZ} give a
method of calculating geometric transformations in a system with many
level crossings and degenerate levels.  Finally we comment on the
system behavior while passing 
through the crossing points. To give a physical picture of the analyzed
problem we apply our results to the process of the three-level Raman
adiabatic passage.

\section{Notion of adiabaticity}\label{adiab}

Consider a quantum system in which we choose a fixed
(parameter-independent) basis of states $\{|1\rangle, |2\rangle,
|3\rangle, \ldots \}$. They can be internal energy states of an atom,
spin pointing "up" and "down" in a fixed spatial direction etc. In general, the
system Hamiltonian $H(\mathbf{R})$ ($\mathbf{R} \in {\cal M}$ is the
set of parameters of the system and ${\cal M}$ the
parameter space, also referred to as the control manifold) contains
terms that couple different states in this basis, and after
diagonalization we obtain the parameter-dependent basis of
instantaneous energy eigenstates 
\begin{equation}
|\bar{m}(\mathbf{R}) \rangle = \sum_n a_{mn}(\mathbf{R}) |n \rangle. 
\label{eq:tva}
\end{equation}
Here $A(\mathbf{R}) = \{a_{mn}(\mathbf{R})\}$ is a unitary matrix
that rotates the fixed basis into the basis of energy
eigenstates. Without any loss of generality we can assume that for
some 
$\mathbf{R}=\mathbf{R}_{diag}$ the matrix $A(\mathbf{R}_{diag})$ is
diagonal, so that at this point the bases are identical. For
$\mathbf{R}_{diag}$ also the 
Hamiltonian written in both bases is diagonal, $H(\mathbf{R}_{diag}) =
H_{e}(\mathbf{R}_{diag}) = \mbox{diag}[E_1(\mathbf{R}_{diag}),
E_2(\mathbf{R}_{diag}),\ldots]$ (the subscript $e$ denotes here the
parameter-dependent basis of the energy eigenstates). For any other
point we can write then
\begin{equation}
H(\mathbf{R}) = A^{\dagger} (\mathbf{R}) H_{e}(\mathbf{R}) A (\mathbf{R}), 
\label{rotatedHa}
\end{equation} 
where $H_{e}$ is (by definition) for any $\mathbf{R}$ a diagonal matrix.
By this construction we see that the information about the energies is
contained only in $H_{e}$, while the information about the states
$|\bar{m}\rangle$ only in $A$ (we shall use this property also later in our
discussion).
In the regions of $\mathbf{R}$ without any level crossing the states
$|\bar{m}\rangle$ are {\em continuous} functions of
$\mathbf{R}$. Indeed, suppose that the system adiabatically follows
$|\bar{m}[\mathbf{R}(t)]\rangle$, and while passing through some
$\mathbf{R}_0=\mathbf{R}(t_0)$, the state changes
discontinuously. Since the states $|\bar{m}\rangle$ form a complete
basis, $|\bar{m}[\mathbf{R}(t_0 - \delta t)]\rangle$
is transformed into $|\bar{m}[\mathbf{R}(t_0 + \delta t)]\rangle =
\sum_{m'} b_{m'} |\bar{m'}[\mathbf{R}(t_0 - \delta t)]\rangle $. But then
this process is equivalent to a discontinuous transition between
different energy levels, which is excluded if the parameters are
varied adiabatically. On the other hand, this observation tells us
that if a level crossing occurs, such discontinuity can be encountered.

\begin{figure}[h] 
\centerline{\resizebox{0.25\textheight}{!}{\rotatebox{0}
{\includegraphics{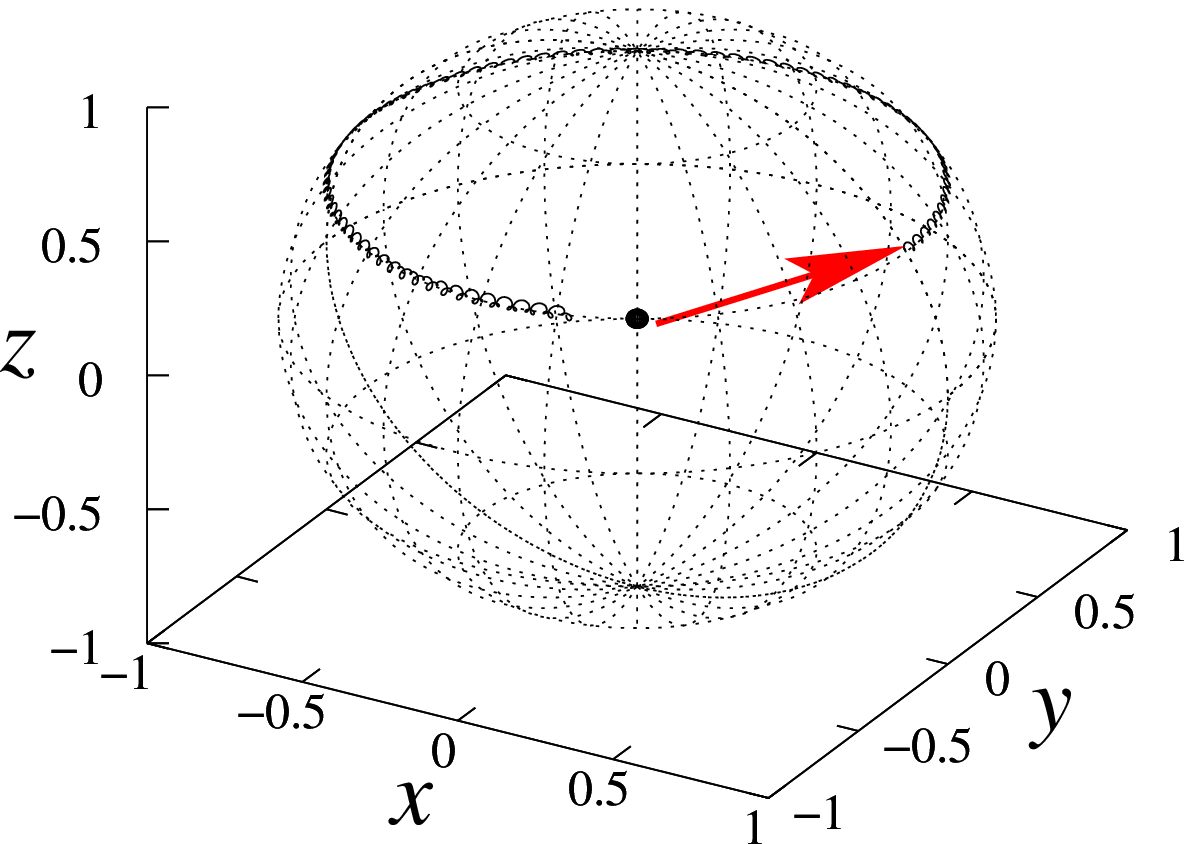}}}}
\centerline{\resizebox{0.25\textheight}{!}{\rotatebox{0}
{\includegraphics{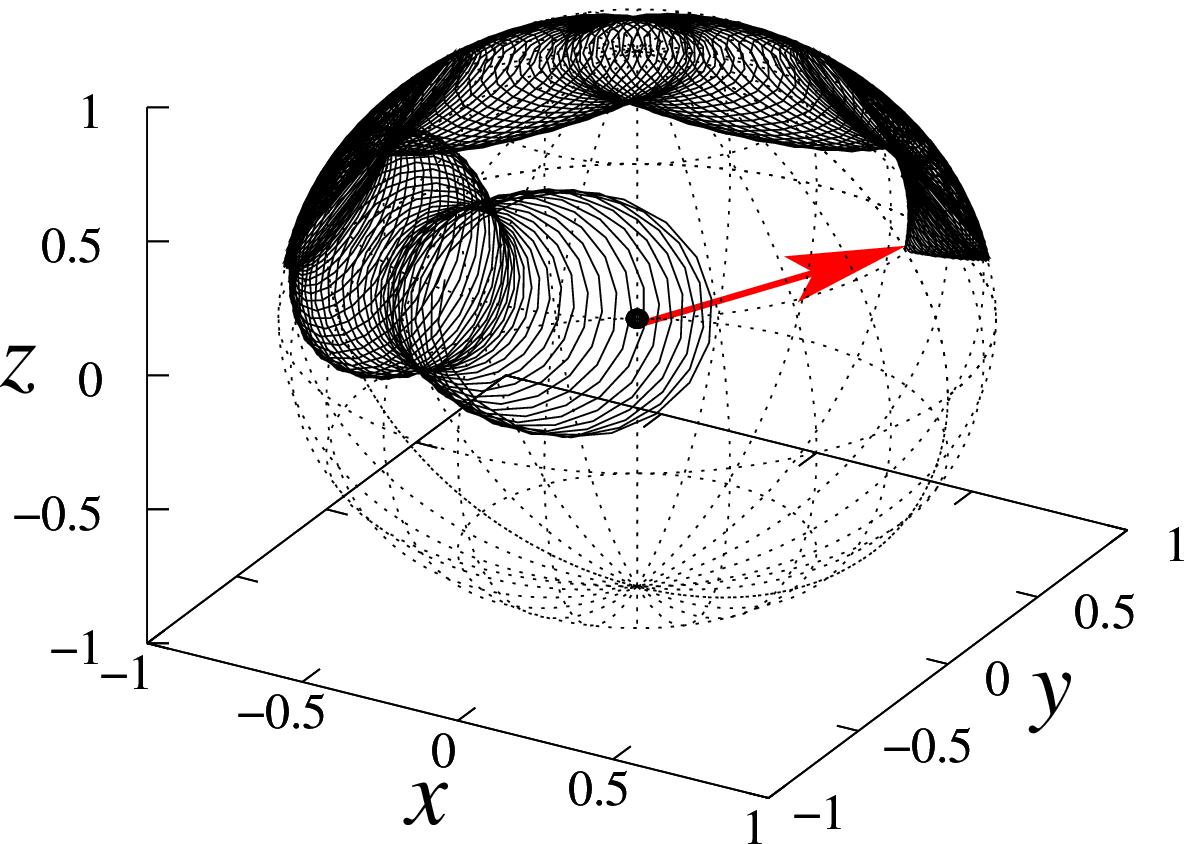}}}}
\centerline{\resizebox{0.25\textheight}{!}{\rotatebox{0}
{\includegraphics{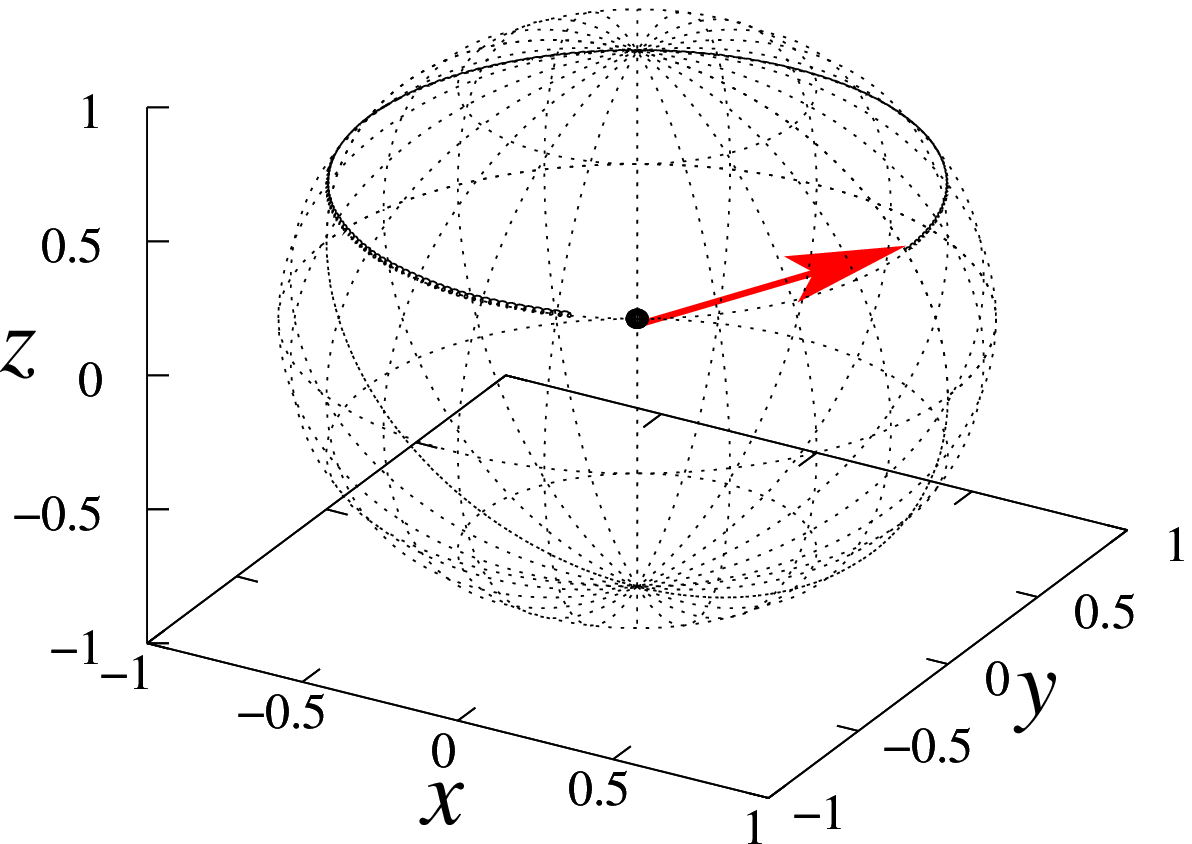}}}}
\caption{(Color online) Illustration of spin-$1/2$ evolution generated
  by variation 
  of external magnetic field. In all cases the angle $\phi $ grows
  linearly with time, slowly (adiabatically) as compared to $B$. The
  variation of the second angle, $\theta $, is different in all cases.
  Upper plot: $\dot{\theta } = 0.005 \times B \cos Bt/2$, middle plot:
  $\dot{\theta } = 0.005 \times B \cos Bt$, lower plot:
  $\dot{\theta } = 0.005 \times B \cos 2Bt$. Since the diagonal terms
  in $\tilde{H}$ are dominated by $B$, the middle plot
  illustrates almost-resonant transition in the rotated frame.}
\label{evol}
\end{figure}

Another problem while approaching the level crossing seems to be the
adiabaticity condition which is hard to satisfy there (and for the
energy separation close to zero implies infinitely slow variation of
the parameters). To analyze this let us first consider two levels that
cross at a point $\mathbf{R}_0$, which for convenience we will shift
to the origin ($\mathbf{R}_0 = 0$). The
subspace of these two states can be characterized by a Hamiltonian of
spin-$1/2$ particle in external magnetic field, for which the energy
eigenstates 
\begin{eqnarray}
\nonumber
|+n \rangle &=& e^{-i \phi} \cos \theta/2 |\uparrow \rangle + \sin
\theta/2 |\downarrow \rangle, \\
|-n \rangle &=& e^{-i (\phi + \pi)} \sin \theta/2 |\uparrow \rangle +
\cos \theta/2 |\downarrow \rangle
\label{spinstates}
\end{eqnarray}
(where $\phi, \theta$ are the spherical angles)
are separated by the energy difference $B$, the strength of the
magnetic field. The level crossing in this geometry corresponds to the
point of zero field. We can now easily find
the adiabaticity condition by going to a frame that rotates together
with the direction of the magnetic
field. The operator that rotates the frame is $A$ [as defined in
Eq.(\ref{eq:tva})], which in this case has the form
\begin{equation}
  A(\theta, \phi) = e^{i \theta /2 \sigma _y} e^{i \phi /2 \sigma _z}.
  \label{eq:hsa}
\end{equation}
The rotated Hamiltonian 
\begin{equation}
  \tilde{H} = i \dot{A} A ^{\dagger} + A H A ^{\dagger},
  \label{eq:0ua}
\end{equation}
for the two-level system reads
\begin{equation}
  \tilde{H} = - {1\over 2} \left(
  \begin{array}{lr}
    B + \dot{\phi }\cos \theta,   & -i \dot{\theta }
    - \dot{\phi } \sin \theta  \\ 
   i \dot{\theta }
    - \dot{\phi } \sin \theta, & - B - \dot{\phi }\cos \theta
  \end{array}
  \right).
  \label{eq:ava}
\end{equation}
The adiabatic theorem states that if the time variation of the
parameters is slow enough, the off-diagonal elements of $\tilde{H}$
are negligible, and so the states precess around the
instantaneous field direction with the frequency $B + \dot{\phi }\cos
\theta$ (the second term gives rise to the Berry phase). It should be
also emphasized (and usually is not) that the Fourier-transformed
off-diagonal terms, $\pm i \dot{\theta }
- \dot{\phi } \sin \theta$ cannot have at $\omega \approx B$ too large
amplitudes as compared to the inverse duration of the process $1/\tau
$ (this is the reason why in the STIRAP experiments, the pulses are
usually chosen to be wide Gaussians). For the purpose of evaluation, this
condition can be written in the form 
\begin{subequations}
\begin{eqnarray}
  f( \omega ) &=& \int_{- \infty }^{\infty }
  dt(\pm i \dot{\theta }
- \dot{\phi } \sin \theta)  \exp \left(-i
  \omega t \right) \\ &&
  \left|\int_{0}^{\tau } dt f( \omega = B(t)) \right| \ll 1,
  \label{eq:bva}
\end{eqnarray} 
\end{subequations}
which together with
\begin{equation}
  \left|B + \dot{\phi } \cos \theta \right|  \gg
  \left|\pm i \dot{\theta } - \dot{\phi } \sin \theta \right|,
  \label{eq:cva}
\end{equation}
gives the adiabatic theorem. A few comments should be made at this
point. First of all, the estimate Eq.~(\ref{eq:bva}) is by no means
less important than Eq.~(\ref{eq:cva}). Indeed, the rotated Hamiltonian
$\tilde{H}$ generates evolution in which the perturbative,
off-diagonal terms may induce transitions between the states parallel
and antiparallel to the direction of the magnetic field. This effect
for different realizations of 
the field variation, but with the same order of magnitude of the
off-diagonal terms in $\tilde{H}$ is illustrated in
Fig.~\ref{evol}. The only difference there is the dominant frequency in the
Fourier transform of these terms. In other words, all the realizations
satisfy well the condition (\ref{eq:cva}), but only two of them
the condition (\ref{eq:bva}), and only in these cases the spin really
follows the field direction. \\
Moreover, in Eq.~(\ref{eq:cva}) the LHS is either
dominated by $B$ 
or $\dot{\phi }$ is too large to allow the condition to hold at all,
and we can omit the term $\dot{\phi } \cos \theta $.
To verify now whether for a given path (which crosses a degeneracy point)
the adiabatic condition can be satisfied let us require the
time dependence of the parameter variation to be constrained by
\begin{equation}
  \left|B\right| \delta =
  \left|\pm i \dot{\theta } - \dot{\phi } \sin \theta \right|,
  \label{eq:jva}
\end{equation}
where $\delta \ll 1$ is a constant, i.e. we want the adiabatic
theorem to be satisfied at each point of the path to the same
extent. The path can be parametrized as 
\begin{equation}
\mathbf{B}(\Gamma ) = \left(B(\Gamma ), \phi
(\Gamma ), \theta (\Gamma )\right).
\label{eq:kva}
\end{equation}
We have the freedom to choose a reference point $\Gamma_0 = 0$ at the
level crossing. The parametrization in Eq.~(\ref{eq:kva}) gives us all the
information about the path along which the parameters vary, and our goal
is to determine the time dependence of $\Gamma $.
Equation (\ref{eq:jva}) can be thus rewritten in the form
\begin{equation}
  \left|B\right| \delta =
  \left|\dot{\Gamma }(\pm i \partial_{\Gamma} \theta -
  \partial_{\Gamma}\phi \sin \theta) \right|.
  \label{eq:mva}
\end{equation}
The solution to this differential equation gives the unique answer to
our question: if $\Gamma (t_0) = 0$ for finite $t_0$, the adiabatic
theorem can be satisfied at each point of the path (down to
the crossing point), otherwise it cannot. 

To illustrate the possible application of this result let us consider
a special class of paths along which the magnetic field in the
spin-$1/2$ scenario is varied. We define $\Gamma = B \tau $, where
$\tau $ is some relevant time scale in the experiment (e.g. its
duration). This guarantees also that $\Gamma _0 = 0$.

\begin{figure}[h] 
\centerline{\resizebox{0.23\textheight}{!}{\rotatebox{0}
{\includegraphics{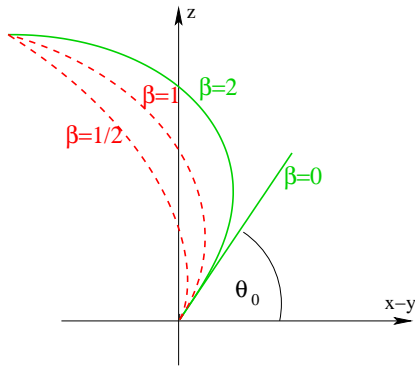}}}}
\caption{(Color online) Realizations of the path $\theta = \theta _0 +
  \Gamma ^{\beta 
  }$, $\phi =$const for various values of $\beta $. Variation along
  the solid paths can be adiabatic at each point, down to the
  degeneracy, while adiabaticity for the
  dashed paths would require infinite time of the process.}
\label{goodpath}
\end{figure}

The class of paths we consider is parametrized by a non-negative, real
$\beta $ by defining 
$\theta = \theta _0 + \Gamma ^{\beta }$ (the degeneracy is approached
from the direction $\theta _0$) and constant $\phi
$. Equation (\ref{eq:mva}) then has the following solutions depending on the 
value of $\beta $:
\begin{equation}
\begin{array}{rl}
 \mbox{adiabatic for any $\Gamma (t)$}, & \beta = 0    \\
 & \\
  \left. \Gamma(t) = \left|{1 - \beta \over \beta
 } {\delta \over \tau } t\right|^{1/(\beta - 1)} \right\} & 0 < \beta
 < 1 \\
 & \\
 \left. \Gamma(t) = \exp \left(\pm {\delta
 \over \beta \tau } t\right) \right\} & \beta = 1 \\
 & \\
 \left. \Gamma(t) = \left|{1 - \beta  \over \beta
 } {\delta \over \tau } t\right|^{1/(\beta - 1)} \right\} & \beta > 1
\end{array}
\label{eq:nva}
\end{equation}
Clearly in the first case the states are time-independent and no
mixing is possible. In the second case, in order to satisfy the
adiabaticity condition we would have to approach the degeneracy
infinitely long. $\beta = 1$ is a critical value, above which the
adiabaticity can indeed be satisfied by choosing the calculated
time dependence of $\Gamma $. The paths for various realizations are
shown in Fig.~\ref{goodpath}.

\section{Adiabatic theorem and geometric transformations with many
  crossing levels}\label{}

One could ask now whether the results are applicable to higher-dimensional
systems, in which some state can cross a degenerate level, two (or
more) degenerate levels cross, or many levels cross at the same
point. Our reasoning is correct in all imaginable cases: we can
consider all pairs of crossing states separately and verify whether 
transitions can be prevented within all of the pairs. At the same time
well-known phenomena hold. For instance,
transitions between states that remain
degenerate during such "insertion" are still possible, and given by
the Wilczek-Zee connection \cite{WZ}.

More systematic explanation makes use of the property
Eq.~(\ref{rotatedHa}). We can define the set of parameters $\mathbf{R}
= \{\{\delta E_{mn}\},\{\theta_k\}\} \in 
{\cal M}$ which has a subset of parameters characterizing the matrix
$H_e$, i.e., the energy differences between the levels. The second
subset contains the angles parametrizing rotations in the Hilbert
space. Using this property we rewrite Eq.~(\ref{rotatedHa}) in a more
general form as
\begin{equation}
H(\mathbf{R}) = A^{\dagger} (\{\theta_k\}) H_{e}(\{\delta E_{mn}\}) A
(\{\theta_k\}). 
\label{separation}
\end{equation} 
Let us now consider two levels, $|\bar{m}\rangle$ and
$|\bar{n}\rangle$ that cross at an isolated point $\mathbf{R}_0 = 0$
(i.e. for all points in the vicinity of $\mathbf{R}_0$ the
parameter $\delta E_{mn}$ is finite). In this setting $\delta E_{mn}$
coincides with the radial component for $\mathbf{R} \rightarrow
0$. This parameter is crucial, as it defines the 
level crossing point, but we need to find all the other relevant parameters
that could give rise to the mixing between the states.
To do this let us consider a point
$\mathbf{R}_1 \in {\cal M}$ which belongs to the path and is shifted
by an infinitesimal vector $d \mathbf {R}$ from $0$ ($d \mathbf {R}$
specifies the direction from which the degeneracy is 
approached). To evaluate the mixing term $\langle \bar{n}(\mathbf{R}_1)
| \nabla |\bar{m}(\mathbf{R}_1) \rangle \cdot d \mathbf{R}$ we make
the expansion $\nabla |\bar{m}(\mathbf{R}_1) \rangle = \sum_{k} b_k
|\bar{k}(\mathbf{R}_1)\rangle$. The only term that can contribute
to the mixing is $b_n |\bar{n}(\mathbf{R}_1)\rangle$, and in the
vicinity of  $\mathbf{R}_1$ the other states can be simply ignored --
indirect mixing, mediated by the other states is possible as the
second (and higher)
order process. The mixing term is, however, proportional to
the first derivative of the states over the parameters and for
sufficiently small 
(in our case infinitesimal) 
variations of the parameters the indirect mixing is
irrelevant. Furthermore, the two states can be conventionally (and
conveniently) parametrized by two spherical angles $\phi _{mn},
\theta _{mn}$, which together with $\delta E_{mn}$ are the only
relevant parameters in our problem. By selecting only three
parameters from the set $\{\{\delta E_{mn}\},\{\theta_k\}\}$ we
reduced the problem again to the 
two-state subspace behavior. Now the angles
parametrizing the states are perpendicular to (independent of) the
radial component of the parameter space $\delta E_{mn}$, and we arrive
again at the 
adiabatic theorem in the form of Eqs.~(\ref{eq:bva}) and (\ref{eq:cva}).

\begin{figure}[h] 
\centerline{\resizebox{0.2\textheight}{!}{\rotatebox{0}
{\includegraphics{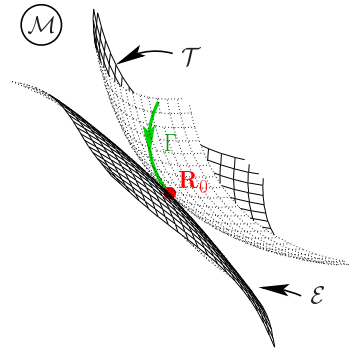}}}}
\caption{(Color online) Schematic structure of the control manifold ${\cal M}$. The
  path $\Gamma $ is formally embedded in the subspace ${\cal T}$
  tangential at $\mathbf{R}_0$ to the degeneracy domain ${\cal E}$.}
\label{fi:manifold}
\end{figure}

It might also happen that due to some symmetries the domain of degeneracy
of the levels $|\bar{m}\rangle$ and $|\bar{n}\rangle$ has a finite
dimensionality. In other words, $\delta E_{mn} = 0$ can define a
subspace ${\cal E} \subset {\cal M}$, where  $\dim {\cal 
  E} = \dim {\cal M} - 1$. Then our simple picture,
in which we identify $\delta E_{mn}$ to be the radial component
clearly fails. However, the path which we traverse  is a
one-dimensional subspace od ${\cal M}$, which close to 
$\mathbf{R}_0$ can be embedded in a subspace ${\cal T} \subset {\cal
  M}$ tangential to 
${\cal E}$ at $\mathbf{R}_0$ (see Fig.~\ref{fi:manifold}). The symmetry
that is present in ${\cal 
  M}$ disappears in ${\cal T}$. Furthermore we can treat ${\cal T}$ as
the actual control manifold -- we formally put only one additional
constraint on the system parameters, which reduces the number of
degrees of freedom by one, but since the path is physically unchanged the
physics of the process is not altered as well. Thus the property
(\ref{rotatedHa}) is still valid and our discussion for the
isolated crossing point is again applicable.

Let us summarize shortly the observations that we have made up to now.
If a quantum system undergoes an adiabatic variation of
parameters $\mathbf{R}$ along a path $\mathbf{R}[\Gamma(t)], t\in
(0;\tau)$ that 
does not contain any level crossing points, all levels are
transformed independently. The resulting transformation
within a {\em $p$-degenerate level} $\{|\bar{m}_1, \ldots, \bar{m}_p\}$
with an energy $E_m$ is given by:
\begin{eqnarray} 
\nonumber
 U_m[\Gamma(\tau ), \Gamma (0)] &=& e^{- i \int_0^{\tau} E_m(t) dt} \\
  \times \lim_{d \tau \rightarrow 
   0} && U_g^p\left\{\mathbf{R}[\Gamma (\tau-d
 \tau)],\mathbf{R}[\Gamma (d \tau)]\right\}. 
\label{result}
\end{eqnarray}
The first exponent on the RHS is the dynamical phase. The
transformation $U_g^p$ is the $p$-dimensional geometric transformation
corresponding to the path $\Gamma$ (the subscript $g$ stands for
``geometric''). It is given by the Wilczek-Zee connection \cite{WZ},
which can be easily derived from Eq.~(\ref{eq:0ua}): within the
degenerate subspace the second term of the Hamiltonian $A H A
^{\dagger}$ is proportional to identity and can be neglected [in
Eq.(\ref{result}) it is included in the dynamical term]. Hence
\begin{eqnarray}
  \nonumber
  \lim_{d \tau \rightarrow 
   0} && U_g^p \left\{\mathbf{R}[\Gamma (\tau-d
 \tau)],\mathbf{R}[\Gamma (d \tau)]\right\}\\ 
\nonumber
= \lim_{d \tau \rightarrow 
   0} && {\cal T} \exp
 \left(\int_{\tau }^{\tau - d\tau } dt \dot{A} A ^{\dagger}\right) \\
  && = {\cal P} \exp \left(\int_{\Gamma } d\mathbf{R} \cdot
  (\nabla_{\mathbf{R}} A)
  A ^{\dagger} 
  \right). 
  \label{eq:bwa}
\end{eqnarray}
Here ${\cal T}$ is the time ordering, and ${\cal P}$ the path-ordering
operator.

Let us comment briefly on this formulation.
The result in Eq.~(\ref{result}) is general, and for distinct levels is
expressed simply by the product of dynamical and geometric
contributions. In such cases the 
ordinary adiabaticity condition holds. If the path begins (ends) at
crossing points we need to examine whether for a given path the
adiabaticity condition can be satisfied, and if it can, evaluate the
connections between (but not exactly at) the initial and the final
point (due to discontinuity of the states). As shown above, if the
path is ``adiabatic,'' the adiabatic
connections [Eq.~(\ref{eq:bwa})] are valid arbitrarily close to the crossing
points. 

The path is parametrized
here by the time $t$ which is the most natural choice. However, the
transformation $U_g^p$ is, as usually, time-independent. It depends only on
the parametric (time) limits of $\mathbf{R}$ at the beginning, and the
end of the path. At the crossing points the discontinuity of the
states $|\bar{m}\rangle$ implies strong dependence of $U_g^p$ on the
direction from which we approach such points.

Finally, for arbitrary adiabatically traversed path in the
parameters space the resulting transformation can be obtained by
dividing the path into pieces that begin and end at crossing points,
evaluating the partial transformations according to Eq.~(\ref{result}),
and eventually multiplying the obtained transformations.

To conclude our analysis let us discuss briefly the nature of
aforementioned discontinuity at the level crossing. It follows clearly
from the degeneracy of the states involved, which gives the freedom to
choose an orthogonal energy eigenbasis. However, if discussed in a
basis of a different observable, its spectrum does not need to be
degenerate at this point. If, in particular, the basis which we
introduced at the beginning -- the parameter-independent basis
$\{|n\rangle\}$ -- is an eigenbasis of an operator that has all
eigenvalues different at the level crossing, amplitudes at the 
states $|n \rangle $ cannot change abruptly. Moreover, the process
under discussion 
is instantaneous (we already know what happens before and after we
reach the crossing point). Probably the most practical indication of
how the system should evolve while passing through the crossing points
give then the conservation laws. So, unless there is some abrupt disturbance
of the system parameters, 
during this infinitesimal time interval quantities that are 
conserved define the ``good'' bases in which amplitudes are continuous.

\section{Nonadiabaticity of the three-level STIRAP}\label{}

\begin{figure}[h] 
 \centerline{\resizebox{0.19\textheight}{!}{\rotatebox{0}
 {\includegraphics{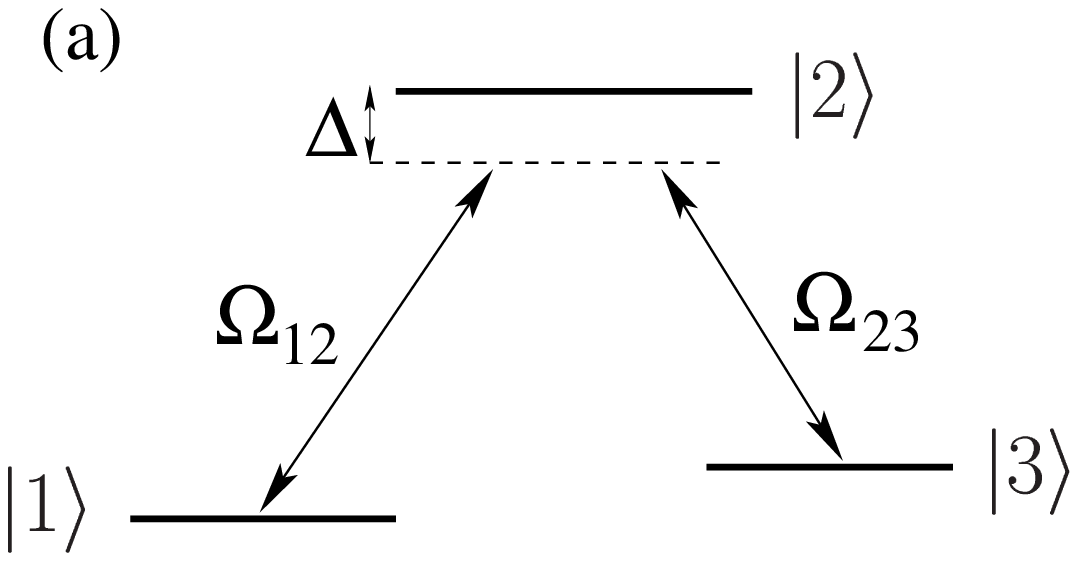}}}}
 \centerline{\resizebox{0.19\textheight}{!}{\rotatebox{0}
 {\includegraphics{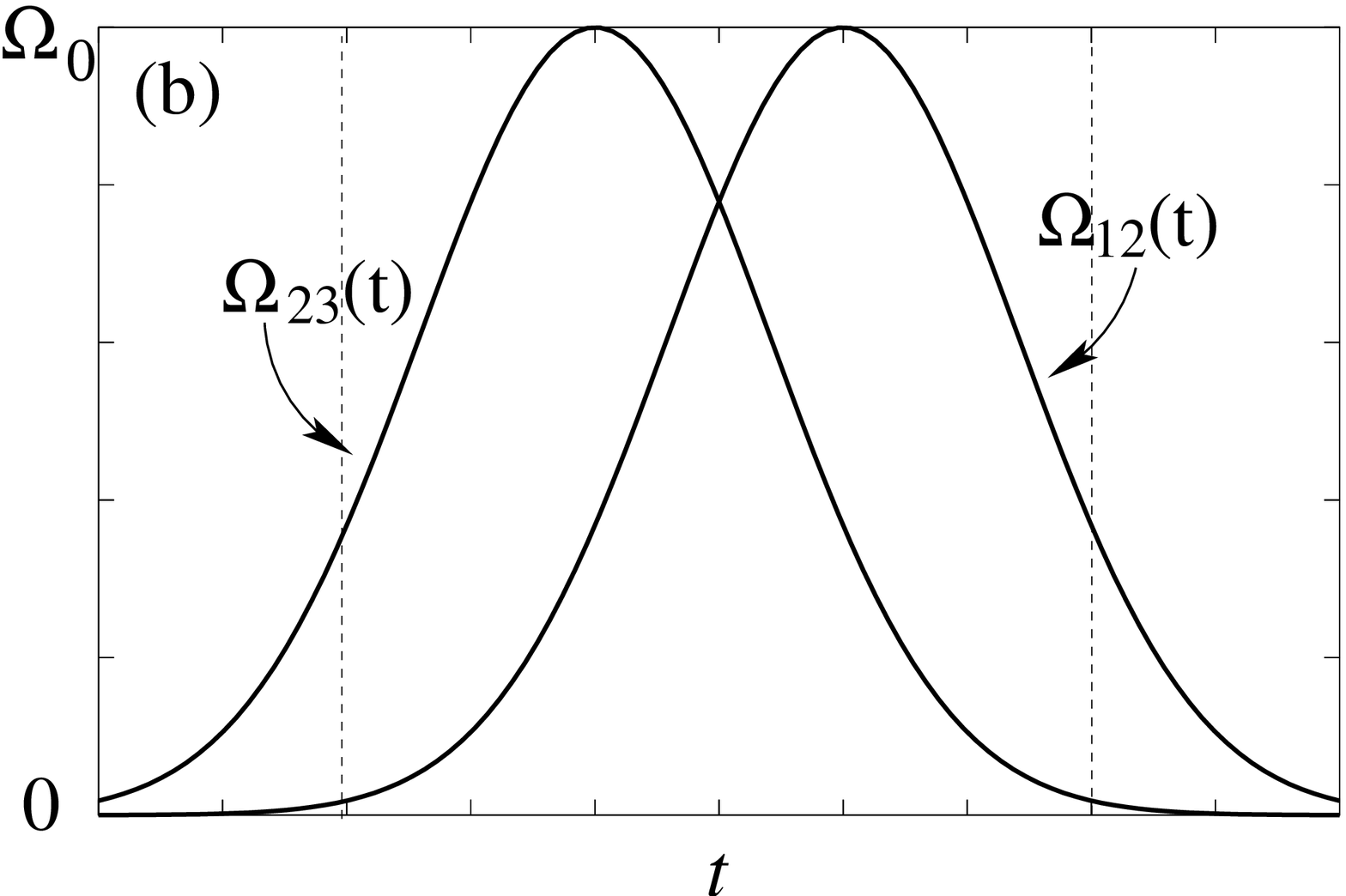}}}}
 \centerline{\resizebox{0.205\textheight}{!}{\rotatebox{0}
 {\includegraphics{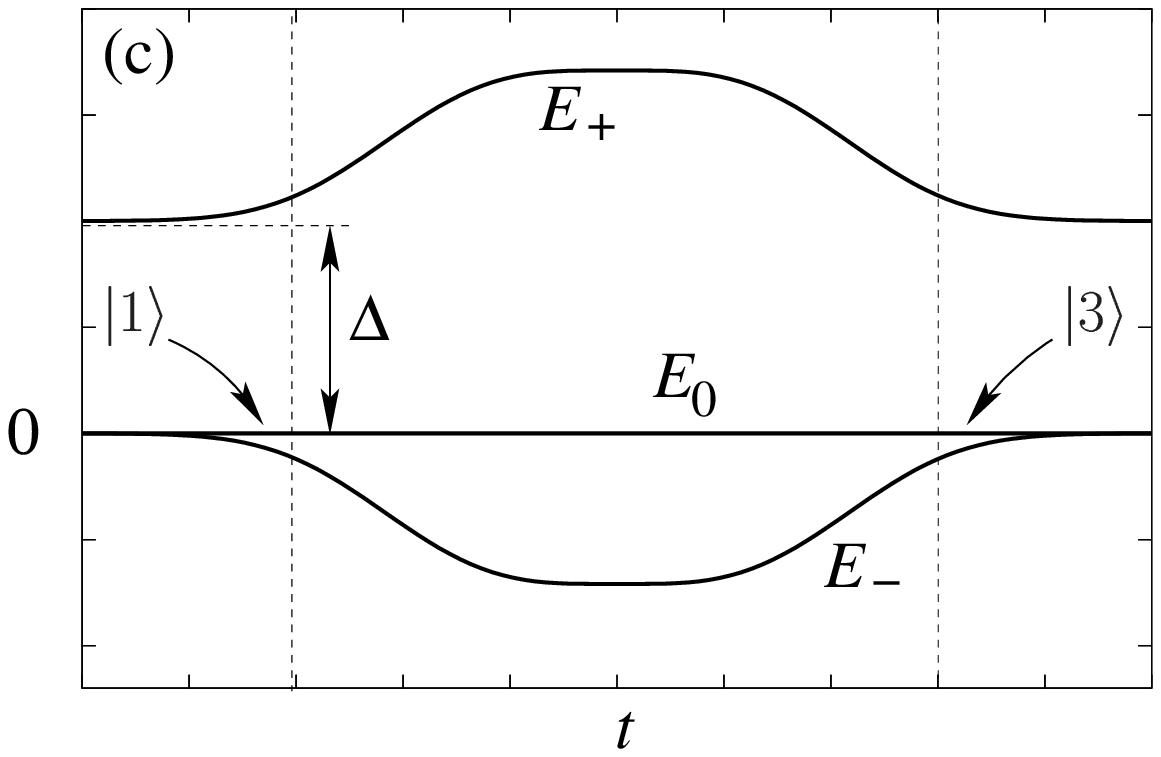}}}}
 \caption{Quantum optical STIRAP process. (a) The level configuration,
   (b) applied pulses, (c) the energy of the eigenstates. The system
   which initially was in the atomic state $|1 \rangle $ follows the
   eigenstate $|E_0 \rangle $ and during the process is transformed
   into $|3 \rangle $.}
 \label{fi:stirap}
 \end{figure}
 \begin{figure}[h] 
 \centerline{\resizebox{0.19\textheight}{!}{\rotatebox{0}
 {\includegraphics{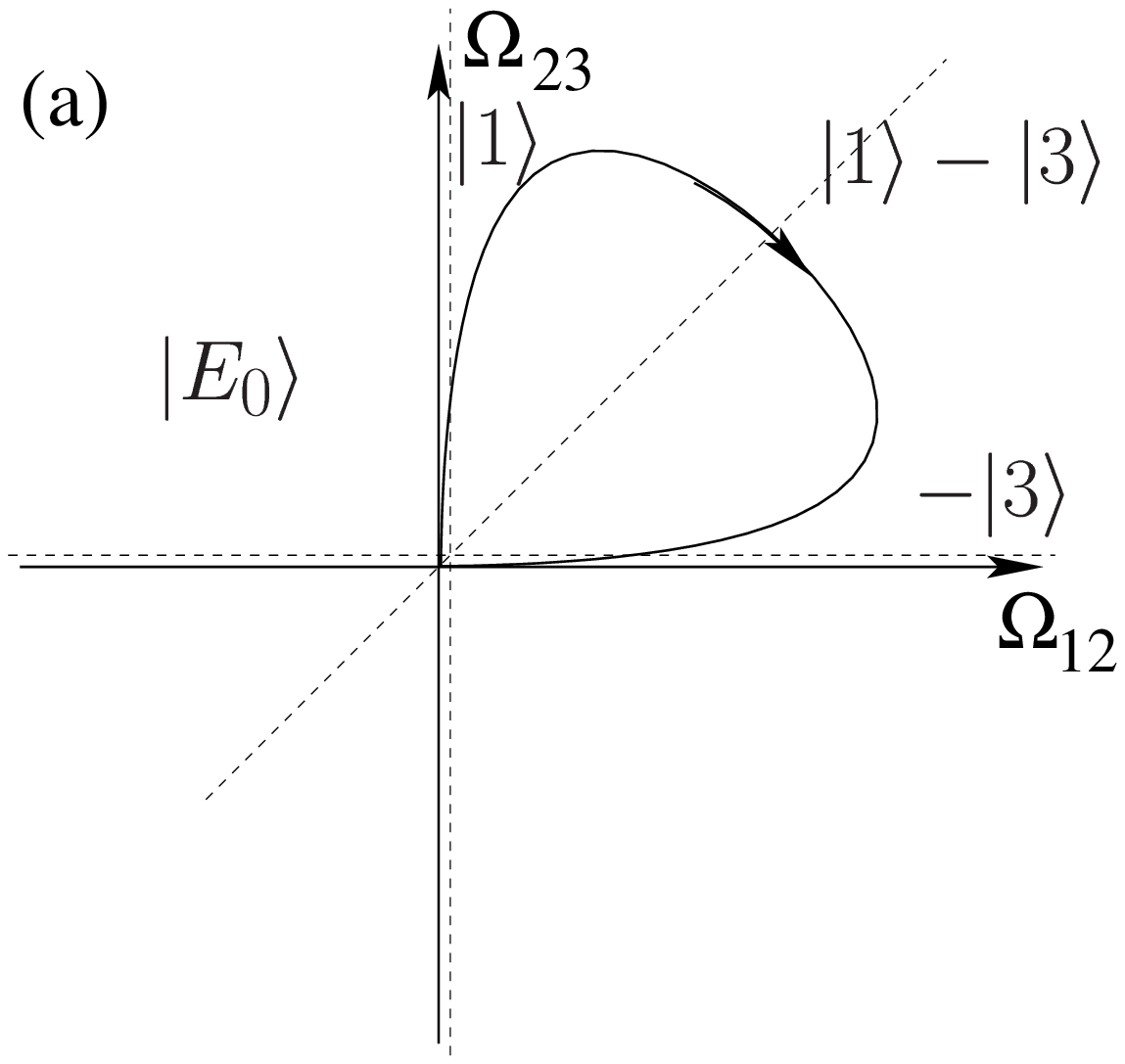}}}}
\vspace{3mm}
 \centerline{\resizebox{0.19\textheight}{!}{\rotatebox{0}
 {\includegraphics{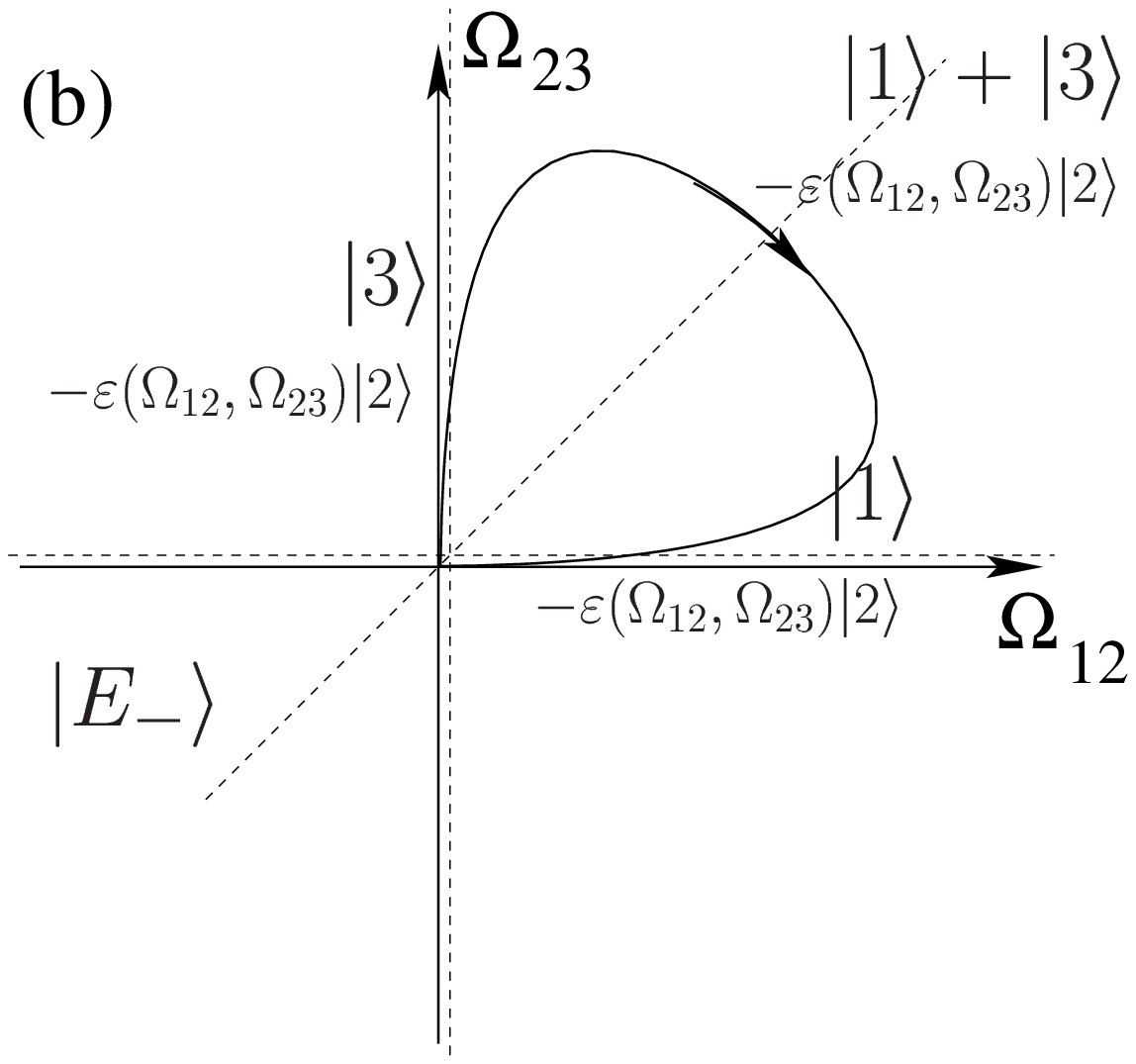}}}}
 \caption{Geometry of the states (a) $|E_0 \rangle $ and (b) $|E_- \rangle
   $. The loops denote the traversed path during the
   passage. Dashed lines are the directions of constant states in
   the 
   subspace $\{|1 \rangle, |3 \rangle \}$. In (b) the amplitude
   $\varepsilon$ changes along these lines, however it does not
   contribute in the first order to the mixing of $|E_0 \rangle $ and
   $|E_- \rangle$.} 
 \label{fi:geome}
\end{figure}
To illustrate how our results apply to real physical systems, let us consider
a three-level atom with internal energy levels $|1\rangle$,
$|2\rangle$, and $|3\rangle$ [see Fig.~\ref{fi:stirap}(a)]. At the initial time
only the state $|1\rangle$ is populated. In order to transfer the
population from the state $|1\rangle$ to $|3\rangle$ we adiabatically
switch on a laser field that nonresonantly couples the states $|2\rangle$
and $|3\rangle$, and then, after some time delay we switch on the
pulse coupling $|1\rangle$ and $|2\rangle$ [see
Fig. \ref{fi:stirap}(b)]. Certainly, since the laser frequency is very close to
the level separation, the process is not adiabatic in the laboratory
frame, but after going over to the rotating frame (in which the amplitudes
absorb the phases resulting from the internal level separation). In
this frame the system 
Hamiltonian (in the basis of the internal levels) has the form
\begin{equation}
H = \left(\begin{array}{ccc}
        0 & - \Omega_{12} & 0 \\
        - \Omega_{12} & \Delta & - \Omega_{23} \\
         0 & - \Omega_{23} & 0
 \end{array}\right),
 \label{true}
\end{equation}
where $\Omega_{12}$, $\Omega_{23}$ are the Rabbi frequencies of the
laser fields, and $\Delta$ is the laser detuning. Diagonalization of the
Hamiltonian yields the energy eigenstates
\begin{eqnarray}
\nonumber
 |E_+\rangle &=& \sin \Theta \sin \Phi |1 \rangle + \cos \Phi |2
 \rangle + \cos \Theta \sin \Phi |3 \rangle, \\ 
 \nonumber
 |E_-\rangle &=& \sin \Theta \cos \Phi |1 \rangle - \sin \Phi |2
 \rangle + \cos \Theta \cos \Phi |3 \rangle, \\ 
 |E_0\rangle &=& \cos \Theta |1 \rangle - \sin \Theta |3 \rangle,
\end{eqnarray}
where
\begin{eqnarray}
 \nonumber
 \Theta &=& \arctan {\Omega_{12} \over \Omega_{23}}, \\
 \Phi &=& {1 \over 2} \arctan \left(2 \sqrt{\Omega_{12}^2+
     \Omega_{23}^2}/\Delta  \right), 
\end{eqnarray}
and the corresponding energies
\begin{equation}
E_{\pm} = {\Delta \over 2} \pm {1 \over 2} \sqrt{4 \Omega_{12}^2 +
   4\Omega_{23}^2+ \Delta^2}, \quad  
 E_0 = 0.
 \label{eq:energ}
\end{equation}
The energies of the levels as functions of time in the usual
experiment are shown in 
Fig.~\ref{fi:stirap}(c). The only states that cross are $|E_0\rangle$ and
$|E_-\rangle $. Some features of their parameter-dependence are shown
in Fig.~\ref{fi:geome}. In particular we see that for the traversed path
the initial and final directions differ and the states are at the
point $(\Omega_{12},\Omega_{23}) = (0, 0)$ discontinuous. Since at
$t=0$ only the state $|1\rangle$ was populated, the system remains (up
to nonadiabatic corrections \cite{stenholm})
during the process in the state $|E_0\rangle $. For the final
direction of the path, the state $|E_0\rangle \equiv |3\rangle $, and
the population is indeed transferred.\\ 

In this setting the only interesting for us energy eigenstates are here the
crossing levels
$|E_0\rangle $ and $|E_- \rangle $ -- transition between this subspace
and the state $|E_+ \rangle $ can be easily suppressed by varying the
parameters on a time scale much longer than $\Delta $ (the usual
adiabatic theorem is applicable). The relevant
quantities in our earlier notation are $B \equiv - E_-$,
$\dot{\phi } = 0$ and $\theta \equiv \Theta $. The paths
used in the STIRAP experiments (see Fig.~\ref{fi:geome}) are parametrized
explicitly by a parameter $\alpha $ in the following way:
\begin{subequations}
\begin{eqnarray}
  \Omega _{12}(\alpha ) &=& \Omega _0 \exp \left[- (\alpha -
  1/2)^2\right],\\
  \Omega _{23}(\alpha ) &=& \Omega _0 \exp \left[-(\alpha + 1/2)^2\right].
  \label{eq:ova}
\end{eqnarray}
\end{subequations}
This parametrization is convenient if $\alpha \propto t$, because of
the evident Gaussian characteristics of the pulses, but this is not
the best choice for our purpose, as the degeneracy point is
approached for $\alpha = \pm \infty $. We will use instead a parameter
$\Gamma $ defined in the following way:
\begin{equation}
  B(\Gamma ) = \Gamma \Delta \delta,
  \label{eq:pva}
\end{equation}
where $\delta \ll 1$ is a constant, and $\Delta $ the laser
detuning. The time $(\Delta \delta )^{-1}$ defines in our case the
adiabatic time scale related to the separation between $|E_+ \rangle $
and the other two levels. To express the RHS of
Eq.~(\ref{eq:mva}) 
in terms of $\Gamma $ we take into consideration the final part of the
path, {\it i.e.} $\alpha \gg 1$. From the ratio $(\Omega _{23}/\Omega
_{12})^2 = \exp(- 4 \alpha )$ we see that $\Omega _{23}$
in Eq.(\ref{eq:energ}) is negligible, and we find $\alpha (\Gamma )$ to be
\begin{equation}
  \alpha (\Gamma ) = {1\over 2} + \sqrt{- {1\over 2} \ln
  \left[{\Delta ^2\over \Omega _0^2}(\Gamma ^2 + \Gamma )\right]}.
  \label{eq:qva}
\end{equation}
The equation of motion for $\Gamma $ now has the following form:
\begin{equation}
  \Gamma \Delta \delta ^2 = |\dot{\Gamma } \partial_{\Gamma }\alpha
  (\Gamma ) / \cosh [2 \alpha (\Gamma )]|
  \label{eq:rva}
\end{equation}
or
\begin{equation}
  \dot{\Gamma } = \pm \Gamma \Delta \delta ^2 \cosh [2 \alpha (\Gamma
  )]/\partial_{\Gamma } \alpha (\Gamma ). 
  \label{eq:sva}
\end{equation}
The sign of the RHS in Eq.~(\ref{eq:sva}) specifies the direction
in which we want to 
move along the loop. We will start at a finite value of
$\Gamma $ and move towards the level
crossing ($\Gamma = 0$), so we choose
the minus sign. 
\begin{figure}[h] 
\centerline{\resizebox{0.3\textheight}{!}{\rotatebox{0}
{\includegraphics{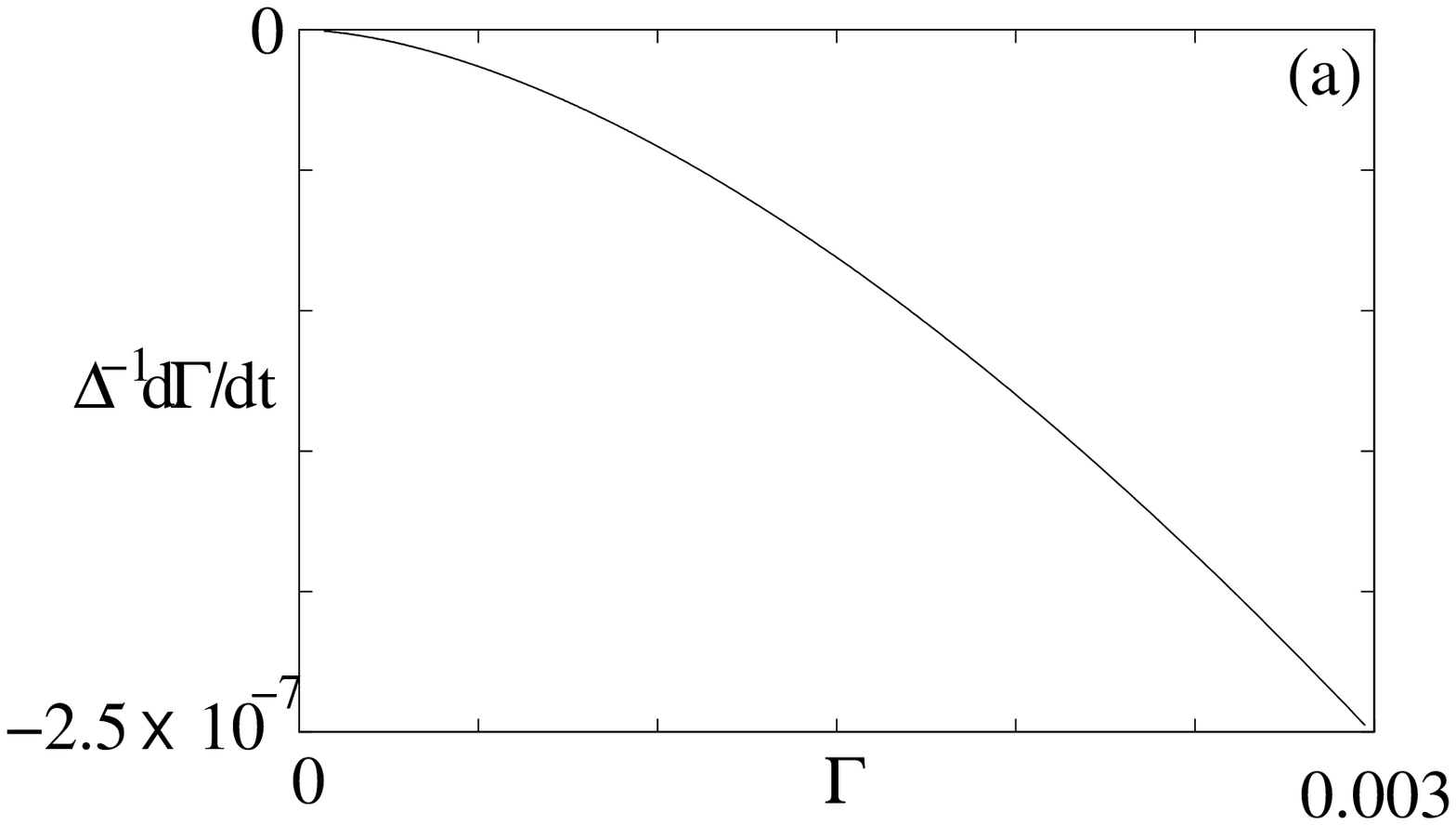}}}}
\centerline{\resizebox{0.3\textheight}{!}{\rotatebox{0}
{\includegraphics{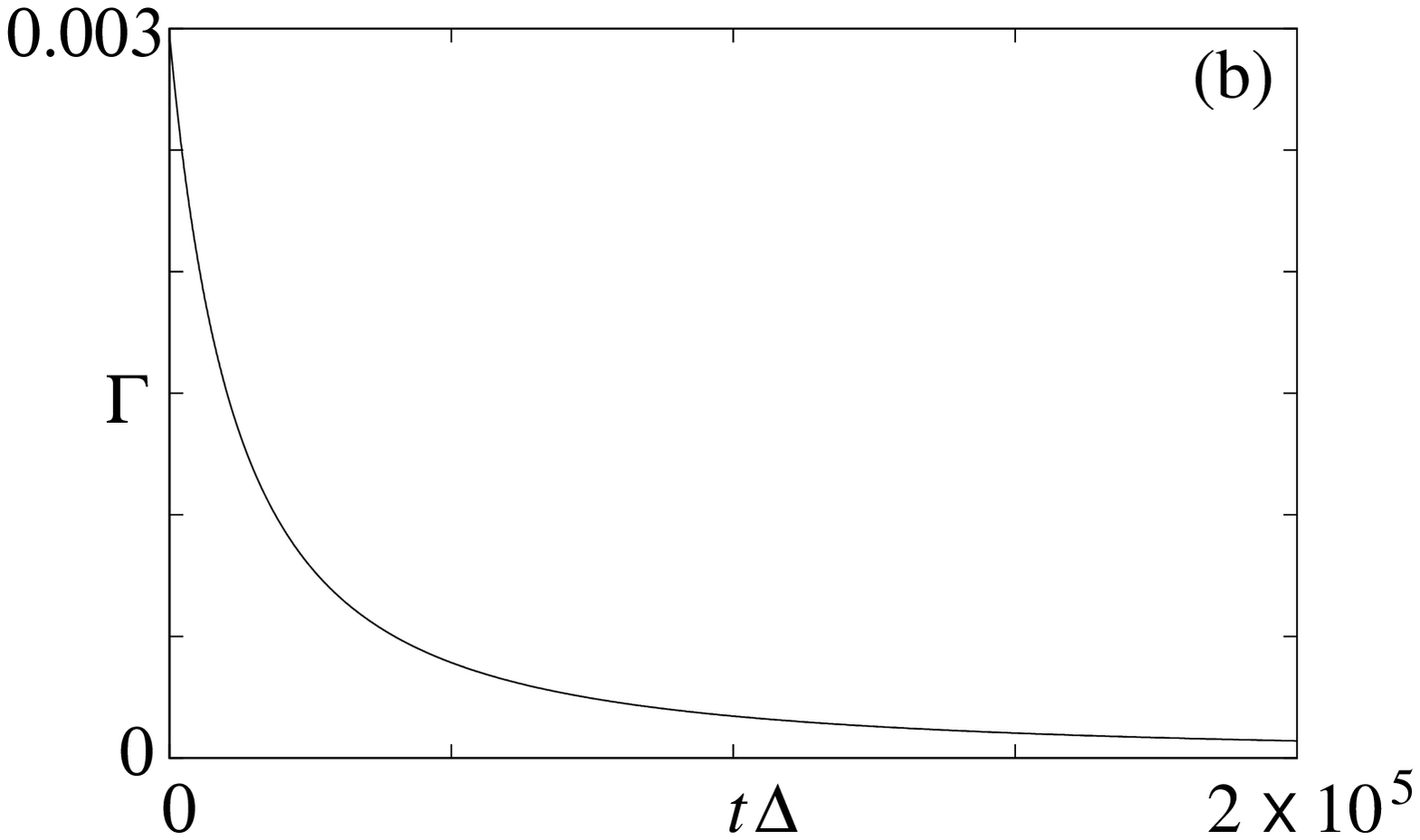}}}}
\centerline{\resizebox{0.3\textheight}{!}{\rotatebox{0}
{\includegraphics{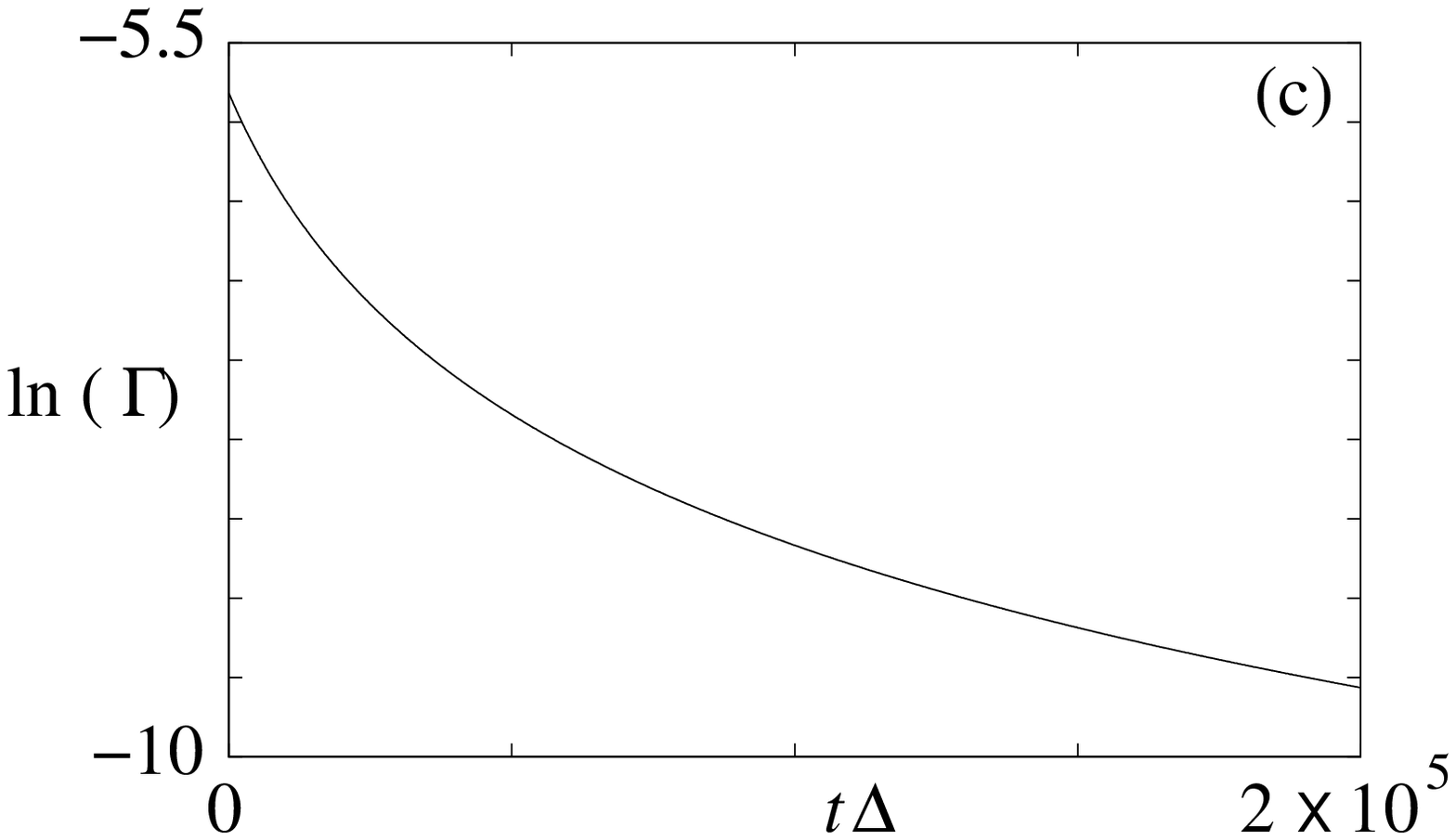}}}}
\caption{Solution of the equation of motion for $\Gamma $,
  parametrizing the path of the three-level STIRAP. (a) $\dot{\Gamma}$
  as a function of $\Gamma $, shows already that near the level 
  crossing, the rate of variation decreases to zero. (b) The solution
  for already a long duration of adiabatic evolution. (c) The
  time dependence of $\ln(\Gamma )$ shows that the parameter
  decreases in time slower than exponentially. In these plots the
  ratio $\Delta/\Omega _0 = 1$, and $\delta = 0.01$.}
\label{fi:stirapath}
\end{figure}
The RHS of Eq.~(\ref{eq:sva}) as the function of $\Gamma $ is shown in
Fig.~\ref{fi:stirapath}(a). Since the rate $\dot{\Gamma }$ reaches $0$ as the
parameter approaches degeneracy, we cannot conclude directly from this
dependence whether the path can be traversed adiabatically or
not. The adiabatic variation of the parameter $\Gamma $
calculated from Eq.~(\ref{eq:sva}) is shown in
Fig.~\ref{fi:stirapath}(b). Apparently we can get arbitrarily close to
the level crossing in an adiabatic manner, but the question
whether we can reach the point exactly in a finite time is answered in
Fig.~\ref{fi:stirapath}(c). Expressing the result in the form $\Gamma
(t)=\exp [f(t)]$ we can see that the function $f(t) = \ln[\Gamma (t)]$
has positive second derivative, which means that $\Gamma (t)$
decreases more slowly than exponentially. Since even at exponential
decrease the degeneracy is reached at $t = \infty $, in our
case the adiabatic variation would last infinitely long as well (and
thus the paths chosen for the STIRAP experiments are not really optimal
for this process, even if the nonadiabatic corrections
\cite{stenholm} turn out to be negligibly small). 

Alternative path for this process,
which would not violate the adiabatic approximation at any point, could be
constructed in the following way: first we switch on the pulse $\Omega
_{23}$ to some finite value $\Omega _0$ while keeping $\Omega _{12} =
0$. Along this 
line the parameters characterizing the states, $\Theta $ and $\Phi $
are constant, and no mixing is possible, regardless of the time of the
variation. Then we could modify the parameters along the
arc $\Omega _{12}^2 + \Omega _{23}^2 = \Omega _0^2$ up to the point
$(\Omega _{12}, \Omega _{23}) = (\Omega _0, 0)$. Along this arc the
energy difference between the states $|E_0 \rangle $ and $|E_- \rangle
$ is constant and equals $B_0 = \sqrt{4 \Omega _0^2 + \Delta
^2}/2 - \Delta /2$. To satisfy the usual adiabatic condition
[Eq.~(\ref{eq:cva})] it suffices to traverse this part of the path in a
time interval $\tau $ much longer than $1/B_0$. However, to make sure
that there will be 
no resonant transitions between the states $|E_0 \rangle $ and $|E_-
\rangle $, the function $\dot{\Theta } =
(\dot{\Omega }_{12} \Omega _{23} - \Omega _{12} \dot{\Omega
}_{23})/\Omega _0^2$ needs to satisfy [according to
  Eq.~(\ref{eq:bva})]
\begin{equation}
  \left|\tau \int_{- \infty }^{\infty } dt \dot{\Theta } e^{- i Bt} \right| \ll 1.
  \label{eq:cwa}
\end{equation}
The path is then closed by switching off the pulse $\Omega _{12}$, again
arbitrarily fast.

\section{Summary}\label{sum}

We have shown that an adiabatic process should not be
understood as a process which is necessarily slow in comparison only
to the energy scales in the problem. Our treatment of the adiabatic
condition proves that for some paths the
transitions between different energy levels can be neglected for a
finite-time evolution in an infinite time scale regions. Moreover,
as many of the other adiabatic phenomena, transformations within one,
nondegenerate level between two
points of degeneracy (corresponding to the same point in the parameter
space) are geometric. The result depends
only on the initial and final directions of the path, not on its details (as
long as it is adiabatic). Our results, together with the theory of
geometric transformations, give also the possibility of designing
experiments which combine the geometric (Berry) phases, holonomic
transformations (within degenerate levels), and adiabatic passage.

\begin{acknowledgments}
The author thanks R. W. Chhajlany, R. Unanyan, and S. Stenholm for
discussions and comments. This work was partly supported by the
DFG-Schwerpunktprogramm ``Quanten-Informationsverarbeitung.'' 
\end{acknowledgments}

\end{document}